# Effect of the crystallographic *c*-axis orientation on the tribological properties of the few-layer PtSe$_2$

Andrii Kozak[1], Michaela Sojkova[2], Filip Gucmann[1,2], Michal Bodík[1,3], Karol Végso[1,3], Edmund Dobrocka[2], Igor Píš[4], Federica Bondino[4], Martin Hulman[2], Peter Šiffalovič[1,3], and Milan Ťapajna[1,2]

[1] Centre for Advanced Materials Application SAS, Dúbravská cesta 9, Bratislava 845 11, Slovakia

[2] Institute of Electrical Engineering SAS, Dúbravská cesta 9, 841 04 Bratislava, Slovakia

[3] Institute of Physics SAS, Dúbravská cesta 9, 845 11 Bratislava, Slovakia

[4] IOM-CNR, Istituto Officina dei Materiali, S.S. 14 km 163.5, 34149 Basovizza, Trieste, Italy

**Abstract**

Two-dimensional (2D) transition metal dichalcogenides are potential candidates for ultrathin solid-state lubricants in low-dimensional systems owing to their flatness, high in-plane mechanical strength, and low shear interlayer strength. Yet, the effects of surface topography and surface chemistry on the tribological properties of 2D layers are still unclear. In this work, we performed a comparative investigation of nanoscale tribological properties of ultra-thin highly-ordered PtSe$_2$ layers deposited on the sapphire substrates with the in-plane and out-of-plane crystallographic orientation of the PtSe$_2$ *c*-axis flakes, and epitaxial PtSe$_2$ layers. PtSe$_2$ *c*-axis orientation was found to has an impact on the nanotribological, morphological and electrical properties of PtSe$_2$, in particular the change in the alignment of the PtSe$_2$ flakes from vertical (VA) to horizontal (HA) led to the lowering of the coefficient of friction from 0.21 to 0.16. This observation was accompanied by an increase in the root-mean-square surface roughness from 1.0 to 1.7 nm for the HA and VA films, respectively. The epitaxial films showed lower friction caused by lowering adhesion when compared to other investigated films, whereas the friction coefficient was similar to films with HA flakes. The observed trends in nanoscale friction is attributed to a different distribution of PtSe$_2$ structure.

**1. Introduction**

Friction at the interfaces of sliding solid-state bodies is related to their physical and chemical characteristics and used environmental conditions [1,2]. In the absence of liquid lubricants between the solid surfaces in contact, the sliding is referred to as dry friction. The dry lubrication is expected to be an advanced solution to reduce the friction and protect the interacting surfaces from mechanical wear in many applications, e.g. in food, textile, and medical industries but also in harsh conditions at extreme temperatures, pressure, or chemically-aggressive environments [3,4]. Dry lubrication is also desirable in the micro- and nanomechanical systems as opposed to the wet lubrication, where the strong capillary forces of the lubricating fluid greatly reduce the efficiency of the moving part of such small systems [5,6]. The wet friction is achieved by introducing the liquid lubricants directly to the system, whereas in the dry sliding, the surface of the solid-state sliding bodies represents the lubricant. To reduce the friction and wear, the protective coatings can be deposited onto the working surface during the system manufacture, or it can be produced as a product of the tribochemical reactions at the surface of the moving parts during the sliding [7].

The layered crystals, self-assembled monolayers, and hard amorphous and crystalline coatings [7–10] were proposed as effective thin-film materials to reduce friction and wear of the individual elements of the low-dimensional systems. Van der Waals (vdW) two-dimensional (2D) layered crystals with a thickness of several atomic layers and extended dimensions in the other two directions became the most popular dry lubricant, owing to the low shear force between their layers [10,11]. Moreover, this structure can exhibit superlubricity by achieving the crystallographic mismatch when different materials or rotation of the layers are used [12,13]. An example of such a structure is a group of transition metal dichalcogenides (TMD). These materials consist of two planes of chalcogen atoms arranged around an interstitial plane of transition metal atoms. There are over 40 different combinations of chalcogen and transition metal atoms [14]. Besides many applications of TMD, including gas sensing, field-effect transistors, energy storage, photodetectors, and catalysis [15–19], these materials have also attracted attention in the tribological studies [8,10]. Similar to graphene [20], they are characterized by atomic flatness, high in-plane mechanical strength, and chemical versatility [21]. All this makes them good candidates for the friction-reducing coatings used in running parts of mechanical devices working at the micro and nanoscale.

Mono- and few-layer graphene, and $MoS_2$ have received a great deal of attention as promising solid-state lubricants [22]. However, a deeper understanding of the friction mechanisms for other 2D material coatings in the widely extended network of layered crystals is still limited [21]. In particular, effects of the surface forces, crystallographic orientation and anisotropy of mechanical properties need to be considered [2,8,10]. A frictional anisotropy of layered crystals attracts attention for a long time due to the difference in the mechanical properties of the crystals in directions parallel and perpendicular to their basal plane [2]. The bulk $MoS_2$ demonstrated an anisotropic coefficient of friction (COF) of 0.1 and 0.26 on the basal and edge surfaces, respectively, originating from its layered structure [2]. It was shown recently that ultrathin $MoS_2$ films demonstrate comparable macroscale friction measured by pin-on-disc with COF of 0.13 and 0.22 on the films with predominant basal and edge $MoS_2$ surfaces, respectively [10]. In contrast, friction at the nanoscale estimated by lateral force microscopy (LFM) showed an order of magnitude smaller friction for the basal-plane films as compared to that of the edge plane. This difference was attributed to weak vdW forces acting along the (001) plane of edge-plane films as well as to their lower surface tension. A theoretical study by Govind Rajan et al [23] suggested the presence of both covalent and partially ionic bonds in the $MoS_2$, where the latter is predicted to cause the electrostatic effects increasing the friction in the edge planes. In addition to the crystal structure, the anisotropic effects can be observed directly in the basal planes of several 2D materials, caused by flexural deformations, lattice commensurability, and stick-slip motion of 2D materials [12,24–26]. This suggests that not only the topographical effect but also structural effects can affect the friction of 2D materials at the nanoscale. The frictional properties of the $PtSe_2$ films were only recently investigated for the first time [27]. However, a comprehensive investigation focusing on the impact of structural properties on the tribological behaviour of $PtSe_2$ layers has not been carried out. In addition, the electronic properties of the $PtSe_2$ layers are sensitive to the strain, which makes them promising for use in optical, mechanical, and electromechanical piezoresistive sensors or in microelectromechanical systems [28–30]. These application fields especially the latter requires detail understanding of the tribological properties of this material.

In this work, we investigated the effect of the crystallographic (001) *c*-axis orientation of few-layer $PtSe_2$ films prepared by selenization of pre-deposited ultra-thin Pt layers (1 and 3 nm thick) on their tribological properties at the nanoscale. The layers were prepared under different conditions. Polycrystalline films with the c-axis perpendicular (1 nm Pt) or parallel (3 nm Pt) to the substrate were formed at a low temperature. The growth at higher temperature resulted in horizontally aligned $PtSe_2$ layers with an epitaxial ordering and different flakes sizes.

The friction and adhesion of the deposited films at nanoscale were characterised by LFM in ambient air measurements. In contrast to our previous experiments on few-layer $MoS_2$ films [10], where we observed a significant difference between friction of the films with different flake orientations with respect to the substrate surface, here the difference in friction for the films with similar grain sizes and different orientation was moderate. In addition, friction measured on the epitaxially-grown samples showed substantially lower friction compared to the non-epitaxial films with different *c*-axis orientation $PtSe_2$ films. This indicates that frictional properties of the $PtSe_2$ films strongly depend on their structure and topography.

## 2. Methodology
### 2.1. Technique of samples preparation

Ultra-thin $PtSe_2$ films with in-plane (also referred to as vertically aligned 'VA' in the following) and out-of-plane (also referred to as horizontally aligned 'HA' in the following) orientation of the *c*-axis of the flakes in respect to the substrate surface, and out-of-plane oriented $PtSe_2$ films consisted of epitaxial grown flakes (referred to as EA in the following) were prepared by selenization process of pre-deposited Pt layers. The Pt layers were deposited on the *c*-plane (0001) sapphire substrates by direct current magnetron sputtering of the platinum target in the Ar atmosphere ($10^{-3}$ mbar) at room temperature. The direct current power and emission current were set to 580 W and 0.18 A, respectively. The Pt thickness ($D_{Pt}$ = 1 and 3 nm) was controlled by the rotation speed of the sample holder. The pre-deposited Pt layers were then selenized in a custom-designed one-zone chemical vapor deposition (CVD) system, where the Pt-coated substrate and Se powder were placed in the same position at the centre of the furnace, so that the substrate and the powder temperature were the same during the selenization process. Keeping the same heating rate of 25°C/min during the selenization, Pt layers deposited on the sapphire substrates were selenized at the temperature of 400°C and 550°C in N$_2$ atmosphere at ambient pressure. All the samples were selenized for 120 minutes.

The low-temperature synthesis resulted in VA and HA films, where the *c*-axis orientation was controlled by changing the thickness of the pre-deposited Pt layers. The VA and HA films with uniformly sized grains were obtained by selenization of 3- and 1-nm thick Pt layer, respectively (Table 1). The high-temperature synthesis provided epitaxial growth of the PtSe$_2$ films on *c*-plane sapphire substrates [31]. The high-temperature samples obtained by selenization of 1- and 3-nm thick Pt layers are referred to as EA$_1$ and EA$_3$, respectively, in the following (Table 1).

**Table 1.** Summary of deposition parameters for synthesis of different PtSe$_2$ films. Lower index for EA samples stands for the thickness of the pre-deposited Pt film.

| Sample reference | VA | HA | EA$_3$ | EA$_1$ |
|---|---|---|---|---|
| Selenization temperature, °C | 400 | 400 | 550 | 550 |
| D$_{Pt}$, nm | 3 | 1 | 3 | 1 |

## 2.2. Orientation of the PtSe$_2$ flakes

The crystallographic orientation of PtSe$_2$ flakes was studied using grazing-incidence wide-angle X-ray scattering (GIWAXS). GIWAXS patterns of the prepared films were measured in air by a custom-designed GIWAXS setup equipped with a Cu Kα microfocus X-ray source (IμS, Incoatec) with the grazing incidence angle of 0.2°. The resulting elastically scattered X-ray signal was detected by a 2D detector (Pilatus 100 K, Dectris).

## 2.3. Structural and chemical analyses of the PtSe$_2$ flakes

The structural properties of the deposited films were analysed by Raman spectroscopy at ambient conditions using Alpha 300R Raman microscope (WiTec, Germany) with the 532 nm excitation wavelength. To avoid the degradation of the ultrathin PtSe$_2$ layers, laser power of 0.5 mW was used. X-ray reflectivity (XRR) measurements were performed using Bruker D8 DISCOVER equipped with a rotating anode (Cu-Kα) working at the power of 12kW.

The chemical state and composition were analyzed by synchrotron radiation X-ray photoelectron spectroscopy (XPS) at BACH beamline at Elettra synchrotron in Trieste, Italy [32]. The beamline was equipped with a Scienta R3000 hemispherical analyzer [33] at an angle of 60° with respect to the X-ray incidence direction. The core-level spectra were measured at a photon energy of 702 eV with a total instrumental resolution of 0.2 eV. The binding energy scale was calibrated to the Au $4f_{7/2}$ peak (84.0 eV) measured on a gold reference. More details on XPS measurements can be found in Refs.[31,34]. Similar to previously published data [31] no elemental platinum or selenium peaks were observed in the spectra. However, the secondary XPS components with a Se $3d_{5/2}$ peak at ~54.7 eV and Pt $4f_{7/2}$ at ~71.8 eV were identified in the low temperature films. This suggests the presence of a second phase identified as non-stoichiometric $PtSe_2$-based phase.

**2.4.    Topography and tribology measurements**

The topography and nanotribological properties of the surface of freshly prepared $PtSe_2$ films were acquired under normal ambient conditions (relative humidity in the range 37 – 47% and temperature 21 ± 1°C) by MultiMode 8 (Bruker, USA) atomic force microscope (AFM). The high-quality topography and frictional maps were achieved with the assistance of an active vibration isolation system. The AFM topography was obtained in tapping mode and was used to determine the root-mean-square (RMS) surface roughness and average grain size ($D_{AFM}$) of all studied samples. For the low-temperature samples with the uniform-sized $PtSe_2$ grains, correlation lengths of the surface features were determined by statistical analysis using 1D or 2D discrete autocorrelation function (ACF) [35]. In general, the ACF reflects the mean distance between the pixels of the same height within the inspected direction and determines the periodicity of the surface features, i.e. $PtSe_2$ grains in this case. Calculated ACFs were approximated by the Gaussian function from which the correlation lengths were obtained. High-temperature $PtSe_2$ layers showed a slightly more textured surface consisting of what appeared to as a mix of in-plane and out-of-plane $PtSe_2$ grains documented by a clear difference in their heights. There, ACF analysis could not have been applied, and manual extraction of grain sizes was done instead. We note that to improve the accuracy of the grain size extraction in these two samples, several different AFM scans were analysed for each of the two samples.

The lateral friction forces were analysed via the LFM in contact mode. In this technique, the AFM tip represents a single asperity that moves over the sample surface in the contact mode. During the move, the torsional deformations of the cantilever are registered. In all AFM experiments, V-shaped silicon nitride cantilevers (Bruker, Scanasyst-Air) with a normal spring constant estimated by thermal noise method of 0.37-0.4 N/m and a nominal tip curvature radius according to probe specification of 2 nm were used. The normal force in the range of 10-50 nN applied during the LFM measurements was defined as a product of the cantilever vertical deflection and its stiffness. As it is rather challenging to accurately calculate the AFM cantilever torsional stiffness, the magnitude of the lateral forces is presented in Volts using averaged values of the forward and backward scanning voltage signals during the LFM.

**2.5. Electrical measurements**

The Hall coefficient and the resistivity were measured by the Van der Pauw method. The charge carrier mobility was then calculated from these quantities. The measurement configuration consists of: i) an electromagnet to set the magnetic field up to 1T by a high power source, ii) a current source Keithley 2400 to set and hold the current on a constant value, and iii) a multimeter Keithley 2700 to measure the resistance and Hall voltage.

**3. Results and discussion**

**3.1. Alignment of the PtSe$_2$ flakes**

The preferred orientation of the crystallographic *c*-axis of the prepared PtSe$_2$ samples with respect to the substrate surfaces was identified using GIWAXS measurements. By tracking the intensity of the distribution of GIWAXS diffraction peaks, the PtSe$_2$ flake alignment with a statistical average over a significant section of the sample surface can be determined. Fig. 1 shows the GIWAXS patterns for the low-temperature PtSe$_2$ films as well as high-temperature films synthesised on the sapphire substrate with different thickness of the pre-deposited Pt layer. GIWAXS pattern with a 001 diffraction peaks at the reciprocal lattice vector length $|q_{001}| \approx 1.15\text{-}1.18$ Å$^{-1}$ [36] are observed for all the films. PtSe$_2$ layers on the sapphire substrate obtained by selenization of the pre-deposited 3-nm thick Pt layer at low temperature showed two symmetrical 001 diffraction spots at $q_{xy} \approx 1.17$ Å$^{-1}$ (Fig. 1a), suggesting the vertical alignment of PtSe$_2$ in respect the substrate surface. In contrast, the samples prepared from the 1 nm Pt layer at low temperature and from 1 and 3 nm Pt layers at high temperatures showed only one 001 spot at $q_z \approx 1.15$ Å$^{-1}$, $q_z \approx 1.18$ Å$^{-1}$, and $q_z \approx 1.18$ Å$^{-1}$, respectively (Fig. 1b-d). Such distribution of diffraction suggests a predominant horizontal alignment of the PtSe$_2$ crystals in respect to the substrate surface. In addition, the patterns with low-temperature films (Fig. 1b) also exhibited faint diffraction rings at the reciprocal lattice vector position close to that of the 001 spots, indicating minor presence of disoriented crystallites in the films. GIWAXS pattern of the epitaxially-grown PtSe$_2$ films are shown in Fig. 1(c) and (d). Samples prepared by selenization of 3-nm thick Pt layers (Fig. 1c), similar to low-temperature HA sample, also showed very faint diffraction ring, suggesting majority of flakes with HA orientation and a negligible contribution of disoriented flakes. Samples prepared from 1-nm thick Pt layer (Fig. 1d) showed only single diffraction spot, suggesting the presence of well-defined HA PtSe$_2$ flakes.

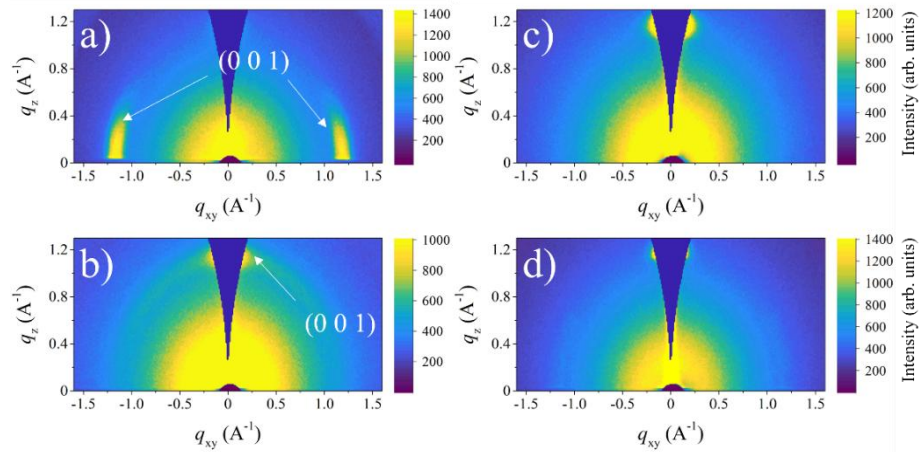

**Fig. 1.** (a, b) GIWAXS patterns of low-temperature PtSe$_2$ films with VA (a) and HA (b) orientation of the *c*-axis. (c, d) GIWAXS patterns of EA PtSe$_2$ films on the sapphire substrate obtained by selenization of predeposited Pt layer with 3- (c) and 1-nm thick (d), respectivelly. A single 001 diffraction spot related to the out-of-plane PtSe$_2$ is partially blocked because it was located in the area of reciprocal space inaccessible to the GIWAXS measurement [37].

### 3.2. Structural analysis

The thickness of as-prepared films was estimated from XRR measurements. Fig. 2 shows XRR pattern of PtSe$_2$ films prepared from 3-nm thick Pt at 400 °C and 550 °C. The oscillations (so-called Kiessig fringes) appear due to the interference of the X-ray beams reflected from the upper and bottom interfaces of the PtSe$_2$ film. The number of oscillations is different suggesting different thickness of the films. This allows calculation of the thickness and roughness ($R_{XRR}$) of the layers. The values of thickness and $R_{XRR}$ of as-prepared films is listed in the Table 2. Note that XRR oscillations can only originate from large-area (mm range) continuous films, while negligible fringes are expected from separate islands of flakes.

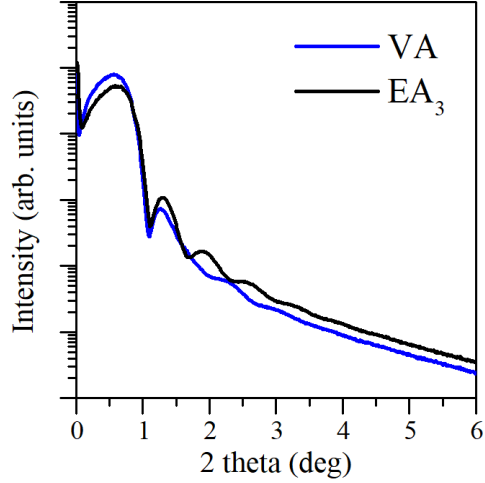

**Fig. 2.** XRR pattern of PtSe$_2$ films prepared from 3 nm thick Pt at 400 and 550 °C.

The structural properties of the synthesized PtSe$_2$ films were further analysed by Raman spectroscopy measurements. The Raman spectra of all deposited films are shown in Fig. 3. Two dominant PtSe$_2$ phonon modes, i.e. A$_{1g}$ (206-208.5 cm$^{-1}$) and E$_g$ (177.5-180.0 cm$^{-1}$) were present in all of the spectra. Additionally, a much less intense longitudinal optical (LO) mode at 234-239 cm$^{-1}$ can be inferred for all the films. Positions of A$_{1g}$ and E$_g$ phonon modes are common in 2D layers with *1T* coordination of PtSe$_2$ [38,39]. The presence of the LO mode suggests the formation of few-layer PtSe$_2$ films [38,40].

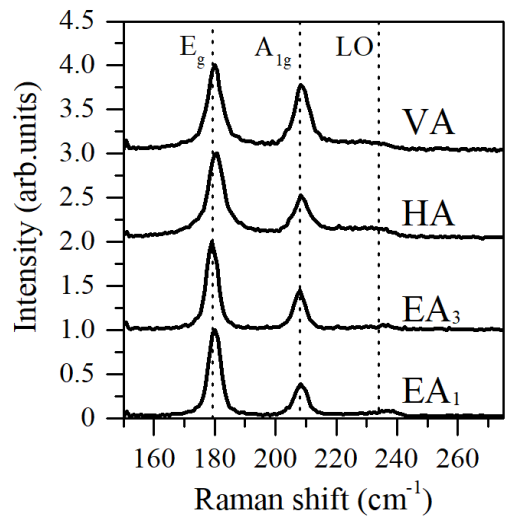

**Fig. 3.** Raman spectra of the PtSe$_2$ films with different orientations, and that of the epitaxial PtSe$_2$ films. All spectra are normalized to the E$_g$ mode.

The spectra of the films prepared by selenization of the 3-nm thick Pt layers are characterized by a redshift of the A$_{1g}$ and E$_g$ modes that is characteristic for increased number of the PtSe$_2$ layers [40,41]. Moreover, the linewidths (FWHM $E_g$, FWHM $A_{1g}$) of the phonon modes of low temperature samples were larger compared to high temperature samples (Table 2). This can indicate a higher density of defects in these samples as well as an increased film thickness [40,41].

Generally, an increase in the $A_{1g}/E_g$ ratio intensity can be associated with an increase in the layer numbers [40,41]. In particular, the $A_{1g}$ mode is more intense on the edges of the flakes [41]. Thus, the lowering of the $A_{1g}$ modes as compared to $E_g$ ones (Table 2) together with the blue shift of $E_g$ mode for the HA PtSe$_2$ films grown at 400°C indicates the lower thickness of the deposited film compared to VA films. In contrast, for high-temperature films redshift of $E_g$ mode and lowering of the A1g mode compared to low-temperature films can indicate lower edge and defects density rather than the increase of the PtSe$_2$ thickness. This hypothesis is further supported by the film thickness XRR measurements and AFM analysis discussed below.

**Table 2.** Structural, tribological and topography-derived parameters of the studied PtSe$_2$ films.

| Sample | VA | HA | EA$_3$ | EA$_1$ |
| --- | --- | --- | --- | --- |
| **Thickness (nm)** | 14 | 4.8 | 12.1 | 4.5 |
| **FWHM $E_g$ (cm$^{-1}$)** | 5.5 | 5.9 | 4.5 | 4.7 |
| **FWHM $A_{1g}$ (cm$^{-1}$)** | 5.7 | 6.0 | 4.6 | 5.1 |
| **$A_{1g} / E_g$** | 0.76 | 0.51 | 0.43 | 0.38 |
| **$R_{RMS}$ (nm)** | 1.7 | 1.0 | 3.9 | 0.8 |
| **$R_{XRR}$ (nm)** | 2.0 | 1.0 | 0.9 | 0.6 |
| **$D_{AFM}$ (nm)** | 10.7 | 6.5 | 105/45* | 63.5/22.5* |
| **COF** | 0.21 | 0.16 | 0.15 | 0.15 |
| **Adhesion (V)** | 0.08 | 0.05 | 0.007 | 0.01 |

$A_{1g}/E_g$ – Raman modes $A_{1g}$ and $E_g$ intensity ratio, $R_{RMS}$ – surface RMS roughness, $D_{AFM}$ – grain size extracted from AFM, * – mean value of the lateral size of the flakes with HA/VA orientation.

### 3.3. Surface topography measurements

The surface topography of the PtSe$_2$ films prepared using different synthesis parameters was investigated by AFM. Fig. 4 shows AFM surface morphology maps of the samples with different orientations of the flakes. It can be inferred that all the films are very smooth and demonstrate a topography typical for polycrystalline materials. The RMS surface roughness of the $PtSe_2$ films is calculated over a scan area of $0.5 \times 0.5$ μm$^2$. The RMS roughness ($R_{RMS}$) of the films decreases from 1.7 to 1.0 nm when the orientation of the films deposited at 400°C changes from VA to HA, respectively. The higher RMS roughness of the VA films compared to HA films deposited at 400 °C is consistent with larger crystalline grains formed in these films. Determined grain sizes ($D_{AFM}$) and $R_{RMS}$ values are summarized in Table 2. As can be inferred from Fig. 4 (a) and (b), low-temperature samples consisted of grains of uniform average size between 5 and 15 nm. Instead, $PtSe_2$ grains observed for high-temperature samples shown in Fig. 4 (c) and (d) were much larger and consisted of two types of grains with different size height. This is likely caused by the presence of larger $PtSe_2$ grains of both alignments in these two samples. We attribute the group of taller grains to the VA flakes, while the larger flat grains to the HA flakes, as there is a similarity between the surface of the VA sample and taller grains in the EA samples. The lateral sizes of grains were estimated to be ~ 20-75 nm for VA and ~ 35-100 nm for HA $PtSe_2$ flakes in sample $EA_1$ and ~42-185 nm for HA and ~12-36 nm for VA $PtSe_2$ flakes in sample $EA_3$. This indicates that the higher temperature during the $PtSe_2$ synthesis facilitates the grain size increase. Simultaneously, different hexagonal flakes in one AFM scan of high-temperature synthesised films are characterised by identical orientation in the basal plane that confirms their epitaxial growth [31].

The increased $R_{RMS}$ (from 0.8 to 3.9 nm) of the films synthesized at 550 °C on the sapphire substrate when the thickness of the pre-deposited Pt layer increased from 1 to 3 nm is related to the increased height of the $PtSe_2$ flakes with vertical orientation. It is interesting to note that $R_{XRR}$ roughness (0.9 nm) for this sample is considerably lower than $R_{RMS}$ (3.9 nm). This is because VA flakes are randomly distributed on the surface in the form of islands and therefore insensitive to XRR measurement. [42]. As the XRR interference takes place only in continuous and uniform films, we can expect that $R_{XRR}$ measured in this specific case corresponds only to part of the film formed by HA $PtSe_2$ flakes in sample $EA_3$. This is consistent with a large scanning area AFM image ($10 \times 10$ μm$^2$, not shown here), where the roughness of the $EA_3$ sample decreased down to 2.5 nm as a result of decreasing in the VA flakes concentration in contrast to high resolution of the $0.5 \times 0.5$ μm$^2$ images.

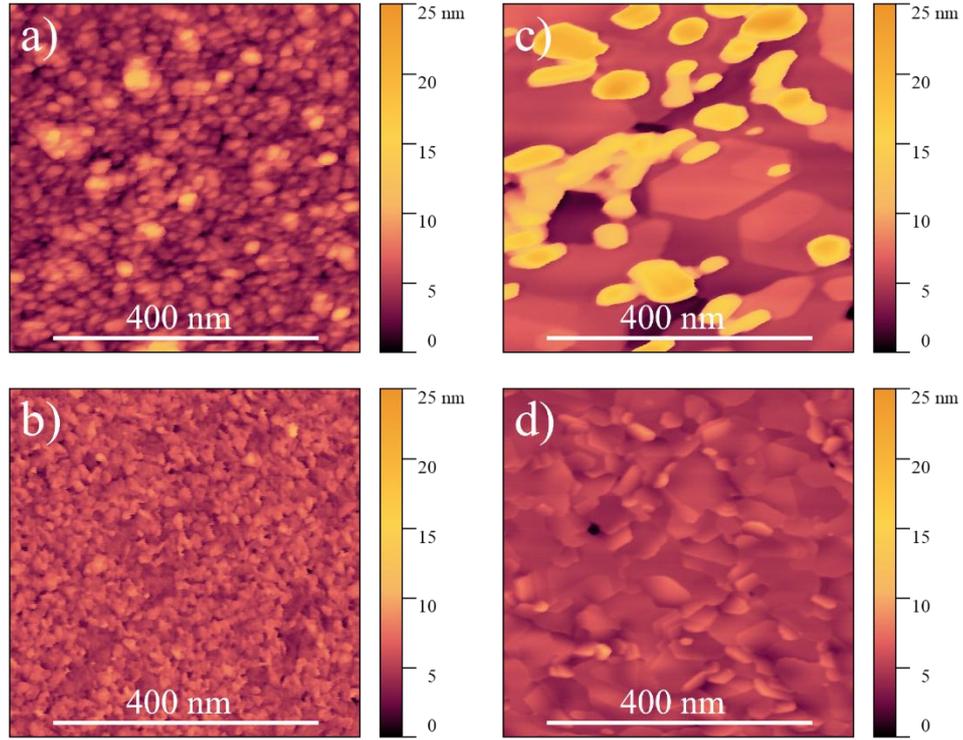

**Fig. 4.** AFM images of PtSe$_2$ films with respective VA and HA orientations of *c*-axis (a) and (b), and that of the epitaxial PtSe$_2$ films (c) and (d) on the sapphire substrates obtained by selenization of 3 and 1 nm Pt layer, respectively.

## 3.4. Friction measurements

LFM measurements were used to determine the nanofrictional properties of PtSe$_2$ flakes. In the LFM imaging, the lateral force between Si$_3$N$_4$ AFM tip and the sample surface is mapped. Figs. 5 (a, b) and (c, d) show LFM maps for the films deposited on the sapphire substrate at 400 and 550°C, respectively. Mean values of the friction were determined from the histograms calculated from the friction maps (exemplified in Fig. 6a for 10 nN applied load) over the entire area of the LFM maps. We note an extremely low value of friction force signal (~0.015 V) for the epitaxial PtSe$_2$ flakes, which was notably lower compared to low-temperature films (~0.124 V and ~0.084 V, for VA and HA films, respectively). Moreover, since no clear friction force dependence on the films thickness was observed here (Table 2) we can exclude the well-known puckering deformation effect and consider the layers as the few-layer films [43][8].

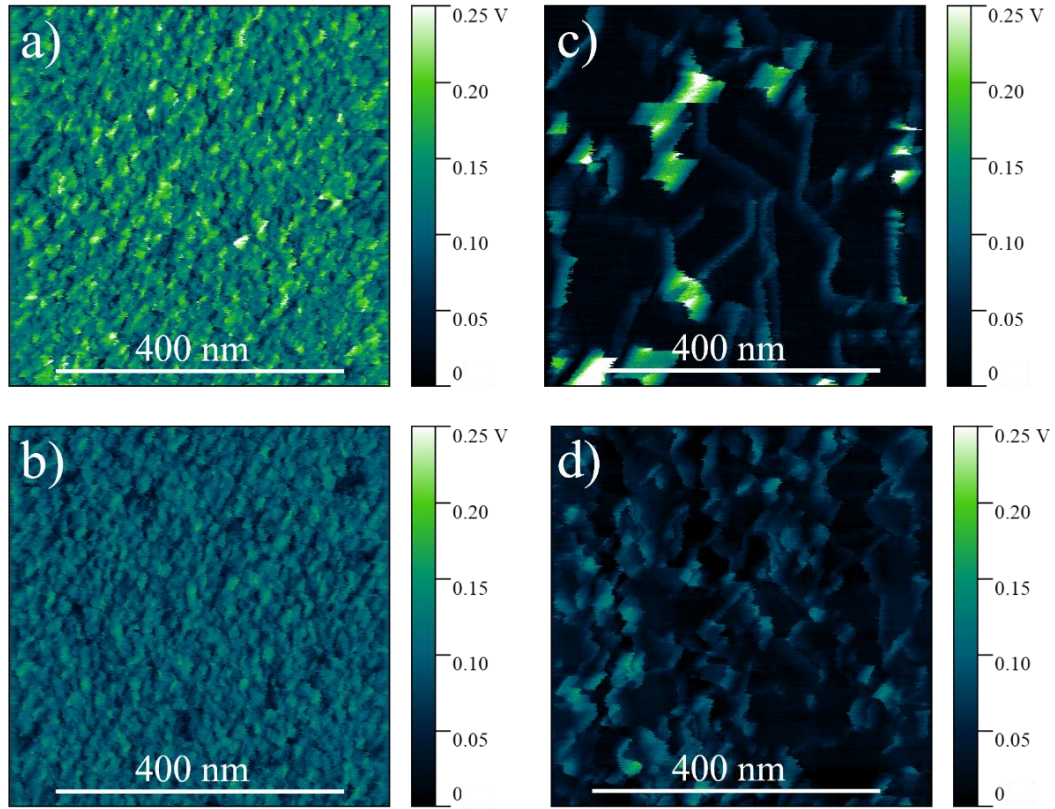

**Fig. 5.** LFM images of low-temperature PtSe$_2$ films with VA (a) and HA (b) orientation of flakes and high-temperature epitaxial PtSe$_2$ films on the sapphire substrate obtained by selenization of 3- (c) and 1-nm thick (d) Pt layer. LFM mapping was performed at 10 nN applied load.

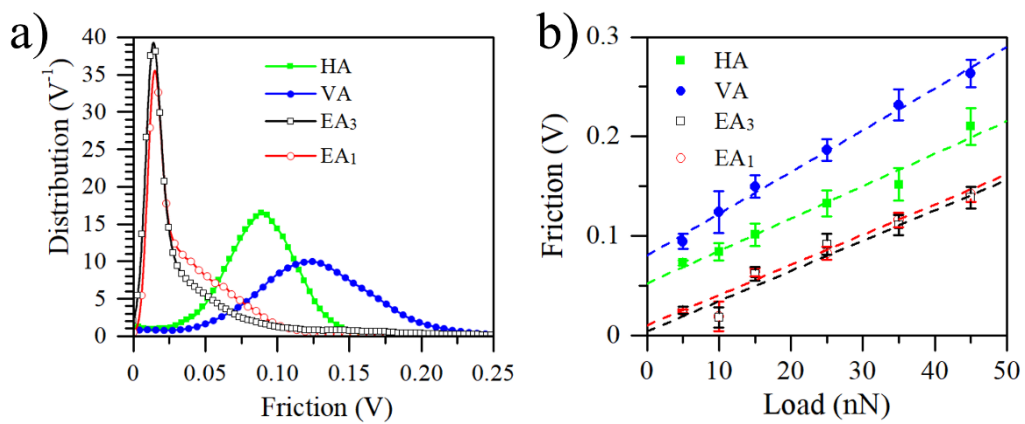

**Fig. 6.** (a) Histogram distributions of lateral signal determined from the LFM images measured at 10 nN showed in Fig. 5. (b) Dependence of the friction signal on the applied load for all the samples determined form the maximum of the LFM hystograms. Error bars represent the standard deviation from Gaussian fitting to the histogram distributions. The dashed lines represent linear fits to the data.

First, let us compare the friction of the low-temperature films with different $PtSe_2$ orientations. The obtained friction signal values showed approximately 30% lower friction of the HA films compared to VA films (0.124 vs 0.084). This behaviour is expected since the higher surface roughness of the VA films compared to the HA ones typically result in a similar trend in the measured friction force, as it also observed at the macroscale. This observation is also consistent with similar measurements on few-layer $MoS_2$ flakes [10]. However, VA and HA $PtSe_2$ films studied here showed very similar values of the friction forces at low applied loads (compared to HA and VA films of $MoS_2$ where one order of magnitude difference in friction was observed [10]), i.e. negligible angular anisotropy of the friction force was observed.

In general, the dry sliding at the micro- and nanoscale is affected by adhesion and deformation components [44]. The adhesion contribution is expected to dominate the friction at the nanoscale, whereas the deformation component prevails at the microscale [45]. Considering the observed small differences in friction signal for samples with different orientation, it is likely that the frictional properties were to some extent affected by the tip-surface interactions. Therefore, the dependences of the friction signal on applied load were measured and are shown in Fig. 6b. In this dependence, the adhesion can be estimated by extrapolation of the linear fit to the data to zero applied load. For low-temperatures films, adhesion signal was found to be slightly lower for HA films as compared to VA films (Table 2). Further, high-temperature EA films show similar adhesion for both Pt thickness, which also is smaller than that determined for the low-temperature samples.

A small difference in the adhesion and friction signal for the films with VA and HA $PtSe_2$ flakes, especially at low applied load, could indicate that the type of the termination of the flake edges (type of atoms or their groups [46]) has an effect on the tribological properties of the films at the nanoscale. In particular, the $PtSe_2$ flakes have *1T* coordination, which results in the formation of the edges with predominant chalcogen atoms termination [47]. In contrast, the *H*-phase metallic edges termination is preferable in $MoS_2$ layers [47]. Therefore, Se atoms termination of the $PtSe_2$ edges and basal plane can produce similar tribological behaviour for the films with different orientations. Also, it was predicted that the $PtSe_2$ layers have entirely covalent bonds [48], leading to the formation of the structures, where any electrostatic effects are precluded; not only in the basal plane but also in the edge planes [23]. The absence of the electrostatic interactions influences the interfacial behaviour of the $PtSe_2$ layers and levels out the interaction energy during the LFM measurements. It is also possible that contamination of the films by hydrocarbons during storage, which saturate dangling bonds and reduce adhesion component, could influence the friction. It is therefore suggested that a higher surface quality of the high-temperature films results in a lower adhesion. In contrast, low-temperature films are more susceptible to surface defects formation due to lower grain sizes, which in turn increase the adhesion and friction forces. The surface defects formation is also consistent with the XPS measurements revealing a traces of non-stochiometric $Pt_xSe_y$ phase.

From the friction force dependence on the applied load (Fig. 6b), we can also estimate the COF of the prepared films by comparing the slope of the fits that of the reference sapphire substrate sample with known values of COF [8]. COFs measured by this way are also listed in Table 2. Note that HA films demonstrate COF almost 1.3 times lower compared to VA films. Further, all HA and EA films showed similar COFs for the films grown at different temperatures, regardless of the crystalline quality. On the other hand, adhesion forces increase in the order EA - HA - VA films. This is in line with our suggestion that the differences in friction forces observed for EA and HA films can be attributed to changes in the adhesion of the films, while changes between low-temperatures films with different orientation could be attributed to changes in adhesion and structural properties (lower friction due to low vdW interaction between layers [10] or oxides on the edges of the flakes can provide easier moving between the grains[2]).

A comparison of the friction behaviour of low-temperature HA and high-temperature $EA_1$ films shows significant differences in the friction signal (c.f. Fig. 6a) and adhesion (Fig. 6b, Table 2). On the other hand, similar COF of 0.16 and 0.15 for HA and $EA_1$ samples, respectively, was extracted for both films. The difference in friction signal is most likely caused by the $PtSe_2$ flakes ordering and crystalline quality. The friction properties are strongly affected by ordering for films with out-of-plane orientation of the c-axis of the PtSe2 flakes in respect to the substrate surface. Due to a higher possibility of epitaxial growth at higher temperatures, the epitaxial films grown on the sapphire substrate films demonstrates lower mosaicity [49], leading to a lower density of defects on the edges and over the basal plane, which also lowers the overall adhesion and friction. Further, similar tribological characteristics of the $EA_1$ and $EA_3$ films indicate that higher crystalline quality resulting from the epitaxial growth has a stronger impact on its tribological properties than the topography or grain size.

### 3.5. Electrical properties of the $PtSe_2$ films

To determine the charge carrier mobility in the $PtSe_2$ films, Hall effect and Van der Pauw measurements were performed using indium contacts cold pressed into the corners of a sample. From these measurements, it is possible to estimate Hall coefficient and charge carrier concentration and afterwards to calculate the charge carrier mobility. The polarity of the Hall coefficient determines the type of conductivity. All as-prepared PtSe$_2$ films showed p-type conductivity. The values of the charge carrier mobility, carrier concentration, and Hall constant are listed in the Table 3. PtSe$_2$ in the bulk form shows semimetallic character and becomes semiconductor after thinning to few layers [50]. Higher values of the mobilities were obtained for thinner films. We suppose that this is caused by rather semimetallic character of thicker PtSe$_2$.

**Table 3.** Electrical properties of PtSe$_2$ films prepared from 1 and 3 nm thick Pt at 400 and 550 °C.

| Sample | VA | HA | EA$_3$ | EA$_1$ |
|---|---|---|---|---|
| Hall mobility (cm$^2$V$^{-1}$s$^{-1}$) | 4.35 | 5.63 | 2.95 | 18.53 |
| Charge carrier concentration (cm$^{-2}$) | $3.51 \times 10^{14}$ | $9.43 \times 10^{13}$ | $1.05 \times 10^{16}$ | $1.25 \times 10^{14}$ |
| Hall constant (cm$^2$/As) | $1.77 \times 10^4$ | $6.6 \times 10^5$ | $5.9 \times 10^5$ | $5 \times 10^5$ |

**Conclusions**

The nanotribological properties of the PtSe$_2$ films prepared by selenization of Pt layers on the sapphire substrates were investigated by the LFM. The friction and adhesion forces of the PtSe$_2$ films were found to be mostly dependent on the structure and quality of the films while the crystallographic orientation of the flakes with respect to the substrate seems to have an impact on the coefficient of friction. The COF was found to be the highest for VA PtSe$_2$ films (0.21), whereas COF in the range of 0.16-0.15 was extracted for HA and EA films. A lower COF for the latter can be attributed to low interlayer interaction. The films grown at 400 °C exhibited similar measured friction force values at low applied load regardless of the orientation, which was attributed to the high adhesion. In contrast, epitaxial films grown at 550 °C with larger crystalline flakes showed a notable decrease in the friction force signal with respect to randomly oriented films as a consequence of increased crystals size. It was also suggested that lower defect densities inside the flakes and on their edges contribute to the lower adhesion and thus friction forces.

**Acknowledgment**

This work was performed during the implementation of the project Building-up Centre for advanced materials application SAS, ITMS project code 313021T081 supported by Research & Innovation Operational Programme funded by the ERDF (50%),Research and Development Agency under the contracts no. APVV-17-0560, APVV-20-0111 and APVV-19-0461, and APVV-19-0365, and Slovak Grant Agency for Science, VEGA 2/0059/21. XPS measurements carried out at BACH beamline of CNR at Elettra synchrotron facility in Trieste (Italy) were performed thanks to the mobility project CNR-SAV-20-03. This project has received funding from the EU-H2020 research and innovation programme under grant agreement No. 654360 having benefitted from the access provided by IOM-CNR in Trieste (Italy) within the framework of the NFFA-Europe Transnational Access Activity. I.P. and F.B. acknowledge funding from EUROFEL project (RoadMap Esfri).